\documentclass[twocolumn,showpacs,preprintnumbers,amsmath,amssymb]{revtex4}

\usepackage{graphicx}
\usepackage{dcolumn}
\usepackage{bm}

\begin{document}

\pacs{78.55Hx, 71.55Ht, 71.70Ch, 77.84Bw}
\title{Optical spectroscopy of trivalent chromium in sol-gel lithium niobate}

\author{J. K. Krebs}
\affiliation{Dept. of Physics and Astronomy, Franklin and Marshall College, Lancaster, PA 17604}%

\author{U. Happek}
\affiliation{ Dept. of Physics and Astronomy, The University of Georgia, Athens, GA 30602-2451}%

\date{\today}

\begin{abstract}
We report on the characterization of sol-gel derived lithium
niobate via  trivalent chromium probe ions, a study that is
motivated by recent reports on the synthesis of high quality
sol-gel lithium niobate (LiNbO$_3$). In order to assess the
quality of sol-gel derived LiNbO$_3$, we incorporate Cr$^{3+}$
during the hydrolysis stage of the sol-gel process. A comparison
of the Cr$^{3+}$ emission and photo-excitation data on both
sol-gel and melt-grown LiNbO$_3$ shows that the sol-gel derived
material is highly stoichiometric.
\end{abstract}

\maketitle

  The need for sol-gel derived ferroelectric  coatings and films for electro-optical
  and piezo-electric applications
   is driving the research for high quality sol-gel lithium niobate (LiNbO$_3$)\cite{Hirano02,Cheng01}.
    Recent reports concentrate
  on optimizing the sol-gel process with the aim
  of producing high quality films
  with preferred orientation. X-ray diffraction and ICP have been the
  primary techniques to assess stoichiometry of
  the samples.  To our knowledge, optical probes, such as
  Cr$^{3+}$ ions, have not been used to characterize the sol-gel
  samples, although this approach has been very successful to show
  fundamental differences between congruent and stoichiometric
  melt-grown LiNbO$_3$ \cite{Salley00}.
  The Cr$^{3+}$ ion is particulary suited to study the quality of LiNbO$_3$,
  because it is arguably the most studied optical
  impurity \cite{Imbusch89}, has a spectrum that is very sensitive
  to the possible lattice sites and disorder in LiNbO$_3$, and can
  be readily incorporated in an early stage of the sol-gel
  process.  In addition, chromium doped lithium niobate  has
stimulated much interest due to the broadband near infrared
luminescence of the chromium ions, with the ultimate goal of
developing a tuneable laser in the visible range through frequency
doubling within the active medium.

 In the current work, we study the optical Cr$^{3+}$ transitions between the $^4A_2$ groundstate and
 the lowest $^4T_2$ and $^2E$ excited electronic states.  These
 transitions are characterized by broad absorption and emission
 bands for the spin allowed transition, and spectrally narrow
 lines for the spin-forbidden transition, respectively. These
 transitions provide information on the stoichiometry and
 disorder of the sol-gel material, and allow a comparison to the detailed study of
 LiNbO$_3$ samples grown from melt.

    For the sample synthesis, we combine 1 M solutions of lithium
    and niobium ethoxides in ethanol such that the ratio of Li:Nb
    is 1:1. An appropriate amount of Cr(NO$_3$)$_3$
    is then dissolved in water (in this study the  Cr:Li ratio was 0.001:1). Under
    constant stirring,  the water-chromium solution is slowly
    added
    \begin{figure}
  \includegraphics[width=3.375in]{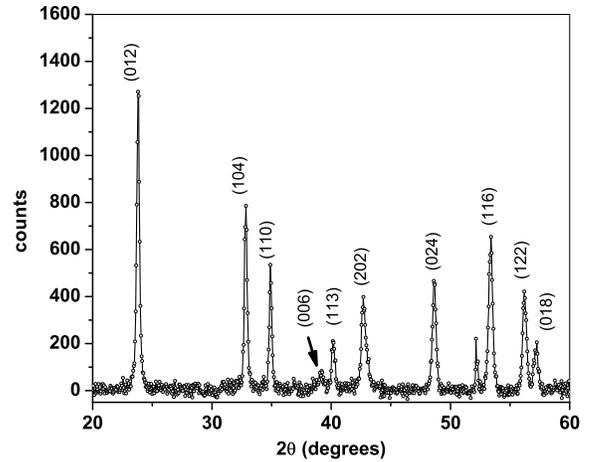}\\
  \caption{X-ray diffraction pattern of sol-gel
LiNbO$_3$:0.1$\%$Cr$^{3+}$.}\label{Figure 1}
\end{figure}
    into the mixed ethoxide. After an initial release of ethanol,
    the solution is capped and aged for several days. The aged solution
    is then dried at room temperature until a fine white powder has
    formed. The powder samples are finally heated in air to 800 $^o$C for
    two hours to remove excess water and organic complexes. The calcined powders
    are characterized by x-ray diffraction utilizing Cu K-alpha
    radiation; a typical pattern is shown in Fig. 1. All observed diffraction peaks
    can be attributed to lithium niobate, and we find no evidence for alternate phases
    such as Li$_{0.88}$H$_{0.12}$NbO$_3$.

    The optical experiments are carried out with the powder sample
    mounted in an Oxford Instruments temperature variable cryostat. For emission
    measurements the sample is excited with a HeNe laser at 632.8 nm. The broadband
    emission was collected with reflective optics and analyzed with
    a Bruker 66v FTIR spectrometer.
     Photoexcitation measurements were performed by mounting the cryostat directly into the sample space of
    a modified Cary 14 spectrophotometer. In both the broadband
    emission and photoexcitation experiments, the emitted light was
    collected through appropriate filters and imaged onto a liquid-nitrogen cooled
    InGaAs detector. For the photoexcitation, a lock-in amplifier was used for signal-to-noise enhancement.
\begin{figure}
  \includegraphics[width=3.375in]{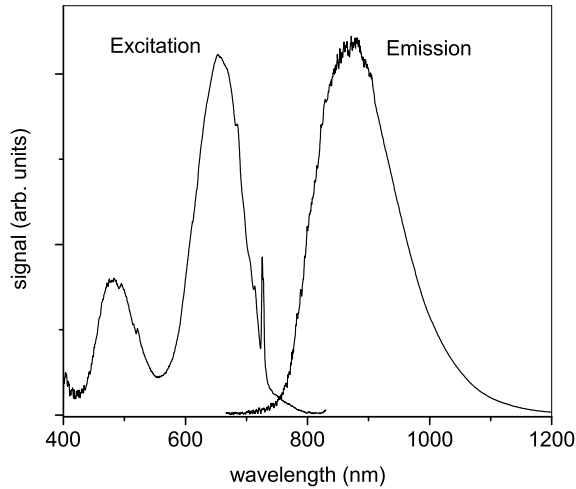}\\
  \caption{Excitation and emission spectra from sol-gel
LiNbO$_3$: 0.1$\% $Cr$^{3+}$.}\label{Figure 2}
\end{figure}

    Experimental results for the sol-gel LiNbO$_3$:0.1\%Cr$^{3+}$ sample are shown in Fig. 2.
    The excitation spectrum (Fig.2, left) observed at $\lambda_{det}\ge$ 900 nm shows the characteristic
    $^4$A$_2$ $\rightarrow$  $^4$T$_2$ and
    $^4$A$_2$ $\rightarrow$ $^4$T$_1$
     broad absorption bands around 655 nm and 480 nm, respectively, as well as the spin-forbidden
     $^4$A$_2$ $\rightarrow$  $^2$E transition around 726 nm.
     The emission spectrum (Fig. 2, right) shows the $^4$T$_2$ $\rightarrow$ $^4$A$_2$ emission
     band under HeNe laser excitation ($\lambda_{ex}$ = 632.8 nm).

    To interpret these results, they are compared to spectra obtained on
    standard melt-grown LiNbO$_3$:Cr$^{3+}$.
Lithium niobate crystals pulled from a melt with equal amounts of
lithium and niobium  crystallize with a Li 3\% deficit
\cite{Byer70}, and are referred to as congruent samples. When
chromium ions are added to these melts, the trivalent impurities
preferentially occupy the crystallographic site of lithium
vacancies, in the following referred to as Cr[Li]. The crystal
field leads to strong absorption into the $^4$T$_2$ and $^4$T$_1$
levels peaking at 654 nm and 478 nm, respectively, giving the
crystals a green color. A shift in the optical spectra is observed
when LiNbO$_3$:Cr$^{3+}$ is codoped with more that 4.5\%
Mg$^{2+}$, resulting in Cr$^{3+}$ impurities preferentially
occupying niobium sites \cite{Macfarlane95}. The resulting change
in the crystal field shifts the $^4$T$_2$ and $^4$T$_1$ absorption
peaks to 715 nm and 540 nm, resulting in a crystal with a red
color. New crystal growth techniques, namely high-temperature
top-seeded solution growth methods (HTTSSG), enabled the
production of LiNbO$_3$ crystals that are stochiometric (i.e. no
Li or Nb deficit)\cite{Polgar97}. These samples produce spectra
with narrower linewidths enabling more precise optical
spectroscopy and better characterization of the various chromium
sites \cite{Salley00}, moreover, the chromium ions in these
stoichiometric melt-grown single crystals are found to occupy both
Li and Nb sites \cite{Salleydiss}.

    In contrast to the Cr[Li] spectrum that dominates in congruent
    LiNbO$_3$ crystals, the sol-gel produced material has a significant contribution
    to the luminescence at excitation wavelengths beyond 750 nm, as evident in a shoulder in the excitation spectrum
    extending from about 750nm to 800 nm in Fig.2.  Fig.3 shows an expanded excitation spectrum in
     this region (dots).
    This shoulder can be  attributed to chromium ions on niobium sites, Cr[Nb].  To quantify
    this contribution to the excitation spectrum, we measured the
    spectra of LiNbO$_3$:Cr$^{3+}$ crystals  with and  without  magnesium codoping (6\%).
\begin{figure}
  \includegraphics[width=3.375in]{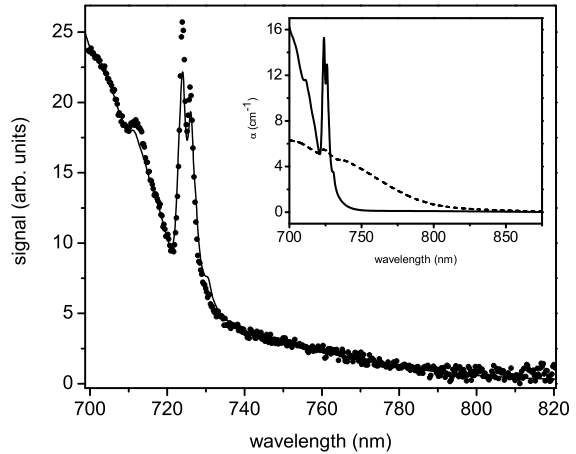}\\
  \caption{Excitation of sol-gel LiNbO$_3$: 0.1$\%
$Cr$^{3+}$ (circles). The inset shows the absorption coefficient
of Cr$^{3+}$ ions in single crystal samples of LiNbO$_3$ and
LiNbO$_3$: 6\% Mg. The fit to the excitation spectrum determines
the 66\% Cr[Li] to 34\% Cr[Nb] ratio.}\label{Figure 3}
\end{figure}

The inset of Fig.3 shows the absorption for the single crystal
    samples of LiNbO$_3$:Cr (solid) and  LiNbO$_3$:Cr:Mg (dashed)
    in the 700 nm to 850 nm range, clearly demonstrating the difference
    of the $^4$T$_2$  onset for the two sites.  The fit to the sol-gel data (Fig. 3, main plot, line) was found by
    varying the contributions to the impurity absorption for each site
    while keeping the total constant, i.e $\alpha_{sol-gel}$ = C$_{Li}\alpha_{Cr[Li]}$+ C$_{Nb}\alpha_{Cr[Nb]}$,
    with C$_{Li}$ + C$_{Nb}$ = 1. The fit shown in Fig.3 (solid line) represents a C$_{Li}$=0.66 contribution
    from the Cr[Li] and a C$_{Nb}$ = 0.34 contribution from Cr[Nb]. This
    approach is justified because our single crystal reference samples
    contained the same amount of chromium ions (0.25\%). A careful
    inspection of  LiNbO$_3$:Cr:Mg (insert Fig.3, dashed line) shows a small peak at 727 nm
    ($^4$A$_2\rightarrow ^2$E transition), due to Cr[Li]. Taking this correction into account, we arrive at
    relative Cr[Li]  and Cr[Nb] concentrations of 0.64 and 0.36, respectively.

The substantial occupation of the niobium site by chromium ions is
in marked difference to melt-grown congruent samples, where the
chromium ions occupy lithium sites exclusively.  In highly
stoichiometric LiNbO$_3$ samples, grown by the HTTSSG method
\cite{Polgar97}, we also find chromium ions on both the lithium
and niobium sites \cite{Salleydiss}, thus our results confirm
reports that the sol-gel process yields highly stoichiometric
samples without the intrinsic Li deficit, which is characteristic
for congruent melt-grown LiNbO$_3$. In stoichiometric single
crystal samples of highest quality, the zero-phonon line of the
$^4$A$_2 \rightarrow ^4$T$_2$ can be observed at low temperature.
In our sample, this transition is broadened, which is indicative
of lattice defects. Our sample also shows the optical signature of
so-called high field sites, which is due to Cr$^{3+}$ ions in a
strong crystal field, resulting in the $^2$E level lying below the
$^4$T$_2$ level, characterized by sharp emission lines at low
temperatures, and which are associated with defects in the
crystal. Thus, while our results indicate that the sol-gel samples
do not have the intrinsic lithium deficit of congruent melt-grown
material, the overall quality of the material does not yet
approach that of the highest quality single crystal material.
However, with the optical absorption and emission spectra of
trivalent chromium being a clear indicator of the crystal quality,
systematic studies are possible to improve the quality of sol-gel
derived LiNbO$_3$.

    In summary, we have shown Cr$^{3+}$ ions are sensitive optical probes of the
     quality of sol-gel derived LiNbO$_3$.   Our experiments confirm that
     sol-gel produced lithium niobate is highly stoichiometric and
     does not contain the intrinsic Li deficit of congruent LiNbO$_3$.
    Our results also show that our sol-gel material contains more defects than
    high quality stoichiometric LiNbO$_3$. With the aid of the chromium optical probe ions, experiments are
    under way to improve the quality of sol-gel LiNbO$_3$ and related ferroelectric
    materials.

\bibliography{jkk05}

\end{document}